\documentclass[twocolumn,showpacs,prl,amsmath,amssymb,epsfig]{revtex4}

\usepackage{graphicx}
\usepackage{dcolumn}
\usepackage{bm}

\begin{document}

\title{A dynamic localization of 2D electrons at mesoscopic length scales}
\author{A. Ghosh, M. Pepper, H. E. Beere, D. A. Ritchie}
\affiliation{Cavendish Laboratory, University of Cambridge,
Madingley Road, Cambridge CB3 0HE, United Kingdom}
\date{\today}

\begin{abstract}
We have investigated the local magneto-transport in high-quality
2D electron systems at low carrier densities. The positive
magneto-resistance in perpendicular magnetic field in the strongly
insulating regime has been measured to evaluate the spatial
concentration of localized states within a mesoscopic region of
the samples. An independent measurement of the electron density
within the same region shows an unexpected correspondence between
the density of electrons in the metallic regime and that of the
localized states in the insulating phase. We have argued that this
correspondence manifests a rigid distribution of electrons at low
densities.
\end{abstract}

 \maketitle

The nature of localization of electrons in low dimensional systems
is a matter of continuing debate. The enhanced effect of Coulomb
interaction in such systems gives rise to quantum many-body ground
states which, depending on the extent of disorder, may display a
wide variety of excitation spectra and real-space distribution of
charge. From extensive theoretical~\cite{gap_th} and
experimental~\cite{gap_expt} investigations, it is now generally
agreed that at zero temperature ($T = 0$), a gap would appear in
the single particle density-of-states (DOS) at the Fermi energy
($E_F$), even though the form of this gap remains controversial.
On the other hand, direct experimental determination of the
spatial distribution of the localized states is significantly
rare. Theoretically, several forms of the distribution have been
predicted in the absence of disorder, which include Wigner
crystals~\cite{tanatar}, bubbles or striped
phases~\cite{bub_strp}. However, when disorder is present, strong
localization is associated with the pinning of electrons to a
static, randomly distributed impurity sites. In this paper, we
have addressed this issue by directly probing the spatial
distribution of localized states within a mesoscopic region of
low-density 2D electron systems (2DES) at various levels of
background disorder. We find that even in the strongly insulating
regime, the areal density of these states ($n_L$) changes with the
electron density ($n_s$), and hence are ``dynamic'' in nature.
When analyzed in the framework of percolation theory, a direct
correspondence between $n_L$ and $n_s$ could be established,
implying an unexpected rigid distribution of the electrons that is
independent of the topology of background disorder, but depends
only on $E_F$ of the system.

In disordered 2DES, the strongly localized regime is identified by
a longitudinal resistivity $\rho \gg h/e^2$, and an activated
(nearest-neighbor) hopping or a variable-range hopping transport,
where $\rho(T) \sim \rho_0\exp{(T_0/T)^p}$, $T_0$ and $p$ being
the relevant energy scale and hopping exponent respectively. In a
weak magnetic field ($B$), applied perpendicular to the plane of
the 2DES, the asymptotic behavior of the wave-function of an
isolated impurity state changes to $\psi(r) \sim \exp{(-r/\xi -
r^3\xi/24\lambda^4)}$, where $\xi$ is the localization length and
$\lambda = \sqrt{\hbar/eB} \gg \xi$~\cite{bs1}. The $B$-induced
compression of the wave-function leads to a strong positive
magneto-resistance (MR), which is expressed analytically as
$\rho(T,B) = \rho(T,B=0)\exp{(\alpha B^2)}$. An expression for the
geometry-dependent factor $\alpha$ has been derived from the shift
in percolation threshold, which in the limit of narrow bandwidth
can be expressed as~\cite{nguen},

\begin{equation}
\label{alpha} \alpha \approx A_L\frac{e^2\xi}{\hbar^2n_L^{3/2}},
\end{equation}

\noindent where $A_L$ is a model-dependent constant of order
unity, whose precise value will be discussed later.
Eq.~\ref{alpha} provides a mechanism of estimating $n_L$ from the
weak-field MR in strongly localized systems, and has been used in
the context of hopping transport in the impurity band of
$\delta$-doped GaAs~\cite{ye}. In the present work we have used
this technique to calculate $n_L$ within a mesoscopic area in
low-disorder 2DES, and compared it to the absolute value of $n_s$
within the same region.

As shown in the schematic of Fig.~1a, mesoscopic segments of
high-mobility Si $\delta$-doped GaAs/AlGaAs heterostructures were
created by the intersection of a narrow wet-etched mesa (width:
$W$) and a transverse Au/NiCr metallic gate (width: $L$). The
doping density $n_\delta = 2.5\times10^{12}$ cm$^{-2}$ was kept
constant for all samples, while the background disorder was tuned
by changing the separation ($\delta_{sp}$) of the dopant layer and
the GaAs-AlGaAs interface. The value of $n_s$ within the active
region depends on the gate voltage ($V_g$) through the specific
capacitance $C_0$. Within the simple capacitor model, $C_0 \simeq
\epsilon_0\epsilon_r/ed_s$, where $d_s$ is the depth of the 2DES
from the surface. For this experiment we have chosen mesoscopic
samples from two heterostructures, which differ in $\delta_{sp}$
and $d_s$, but identical in all other structural and geometrical
aspects. The relevant parameters are given in Table~\ref{tab1}.
Note that sample A07 is more disordered than A78 due to a smaller
$\delta_{sp}$, and hence has lower electron mobility ($\mu$). All
experiments were performed at $T \approx 0.3$~K unless specified
otherwise. In order to minimize the effect of contact and series
resistances, the electrical measurements were carried out only in
4-probe geometry with a low-frequency ($\sim 7.2$~Hz) ac
excitation current of $\sim 0.01 - 0.1$ nA. The maximum measured
resistance was limited to $\sim 1.5$ M$\Omega$ to avoid any
systematic error arising from stray capacitances.

The first part of the experiment evaluates the $V_g$$-$dependence
of $n_s$ when $n_s$ is sufficiently high such that $\rho \ll
h/e^2$ and quantum Hall-based methods are applicable. However, due
to the restricted dimensions of the active region, in particular
the smallness of $L$, conventional Hall or Shubnikov-de Haas
measurements are experimentally unfeasible in these samples.
Hence, we have used an alternative technique based on the
reflection of edge-states, which was originally developed to study
backscattering of edge-states by well-defined potential
barriers~\cite{haug}. The technique is schematically represented
in Fig.~1a. Briefly, when the perpendicular $B$ corresponds to
$\nu_0$ edge channels in the ungated part of the Hall-bar, a
negative $V_g$ will cause $\nu_0 - \nu$ of the channels to get
reflected at the gate, where $\nu$ is the filling factor within
the active region. As a function of $V_g$, this leads to plateaus
in the four-probe resistance ($R_{12,34}$) at $R_{12,34} =
(h/e^2)(1/\nu - 1/\nu_0)$, when the contacts (1-4) in Fig.~1a are
close to ideal with no reflection and perfect transmission. This
is illustrated in Fig.~1b, where the $V_g$$-$dependence of
$R_{12,34}$ is measured in sample A78 at a constant $B$
corresponding to $\nu_0 = 10$. Three plateaus corresponding to
$\nu =$ 8, 6 and 4 appear nearly exactly at the theoretically
expected values (horizontal dashed lines). The $V_g$ at the center
of these plateaus then corresponds to an electron density of $n_s
= eB\nu/h$, while the width of the plateau represents the maximum
uncertainty. For both samples, we have repeated this measurement
for different values of $\nu_0$ (within spin degeneracy).

\begin{table}
\caption{\label{tab1} Relevant parameters of the samples used.
$\mu(0)$ and $n_s(0)$ are as-grown mobility and electron density
respectively.}
\begin{ruledtabular}
\begin{tabular}{lllllll}
Sample&$\mu(0)$&$n_s(0)$&$\delta_{sp}$&$d_s$&$L\times W$&$C_0$\\
&cm$^2$/V-s&cm$^{-2}$&nm&nm&$\mu$m$\times
\mu$m&cm$^{-2}$V$^{-1}$\\\hline
A07&$0.6\times10^6$&$2.9\times10^{11}$&20&120&$2\times8$&$57\times10^{10}$\\
A78&$1.8\times10^6$&$2.1\times10^{11}$&40&290&$2\times8$&$24\times10^{10}$\\
\end{tabular}
\end{ruledtabular}
\end{table}

Fig.~1c shows $n_s$ as a function of $V_g$ for samples A07 (empty
symbols) and A78 (filled symbols). An advantage of this method is
the negligible uncertainty in the $y$-direction since $n_s$ is
derived from a known integer ($\nu$). At a sufficiently negative
$V_g$ both samples display the expected linear dependence of $n_s$
on $V_g$. In A07, the deviation from linearity at higher $V_g$ can
be explained from screening of $V_g$ by accumulation of carriers
in the dopant layer~\cite{hirakawa}. When the linear part is
fitted with $n_s = C_s(V_g - V_{gs}^0)$, the slope $C_s$ was found
to be $\approx 55.5\times10^{10}$ cm$^{-2}$V$^{-1}$ for A07 and
$\approx 23.0\times10^{10}$ cm$^{-2}$V$^{-1}$ for A78, agreeing
satisfactorily with corresponding values of $C_0$. The intercept
$V_{gs}^0$ was found to be $V_{gs}^0 = -0.917\pm0.004$ V and
$-0.804\pm0.002$ V for A07 and A78 respectively.

We now consider the strongly localized regime, where $\rho \gg
h/e^2$. As shown in the insets of Fig.~2, at $B = 0$, this regime
corresponds to $V_g \lesssim -0.86$ V for A07 and $V_g \lesssim
-0.74$ V for A78. Increasing $B$ from zero initially results in
some negative MR in both samples, followed by an exponential rise
in $\rho(B)$ at higher $B$. The dependence of $\ln{\rho}$ on $B^2$
at different $V_g$ for both samples is shown in Fig.~2. In order
to ensure consistency, the analysis is limited to the regime of
$V_g$ for which $\rho(B=0) \gtrsim 2h/e^2$. From Fig.~2, within
the measured range of $\rho$, $\ln{\rho}$ varies linearly with
$B^2$ over a field range of 1.2 T $\lesssim B \lesssim$ 3 T for
A07 and 0.5 T $\lesssim B \lesssim$ 1.3 T for A78. The lower limit
of $B$ is determined by the extent of negative MR, which arises
from an interference among the hopping paths~\cite{int_hop}. The
upper cut-off in $B$ probably arises from the strong-field
condition $r \ll \lambda^2/\xi$ for the asymptotic form the
wave-function $\psi(r)$~\cite{bs1}. This is not fully understood
at present and will be discussed in a future publication.

The most striking aspect of the data shown in Fig.~2 is the
continuous decrease in the slope $\alpha$ of $\ln{\rho} - B^2$
traces as $V_g$ is made more positive (i.e. as $n_s$ is
increased). This is observed in both samples (Fig.~3). Note that
from the expression for $\alpha$ in conventional hopping framework
(Eq.~\ref{alpha}), $\alpha$ is expected to increase with
increasing $V_g$ since the localization length $\xi$ increases
rapidly as the system becomes more metallic ($n_L$ remains
constant since it is assumed to arise from frozen impurity
distribution). This has been confirmed in hopping MR in the Na$^+$
impurity band of Si-MOSFET's, where $\alpha$ was found to increase
as the Fermi energy was swept towards band
half-filling~\cite{fowler}. In our case, we also note that the
absolute magnitude of $\alpha$ is about factor of $\sim$ 2 larger
in A78 than A07 over the same range of $\rho$.

If we assume the validity of Eq.~\ref{alpha}, the
counter-intuitive trend of $\alpha$ in Fig.~3 forces $n_L$ to be
$V_g$-dependent. Following Ref.~\cite{ye}, since $\rho \gg h/e^2$,
we first replace $\xi$ of the hydrogenic wave-function of the
strongly localized states by $a_B^*$, where $a_B^*$ is the
effective Bohr radius of electrons. However, an evaluation of
$n_L$ from Eq.~\ref{alpha} also requires the magnitude of $A_L$.
If the system consists of only two isolated localized states, one
obtains $A_L$ simply from the overlap integral as $A_L = 1/12
\approx 0.083$~\cite{sav}. However, in the hopping regime there
are a large number of states, and the relation $A_L =
(N_c/\pi)^{3/2}/12$ can be obtained following Nguen's analysis of
the $B$-induced shift in percolation threshold. Here $N_c$ is the
dimensionality-dependent number of bonds per site in the random
resistor network~\cite{nguen}. $N_c = 4.5$ or 2.7, for zero or
finite width of the impurity band with respect to
$k_BT$~\cite{fowler,hayden}. This gives $A_L \approx 0.143$ and
0.0664 respectively. Note that all the estimates of $A_L$ are well
within the order of unity, and hence only has a fine tuning effect
on the absolute magnitude of $n_L$. In the subsequent analysis we
have used $A_L = 0.0664$, even though the effect of other values
of $A_L$ has also been discussed.

Fig.~4a shows the $V_g$$-$dependence of $n_L$ evaluated using
Eq.~\ref{alpha} and the experimentally obtained $\alpha$. Note
that for both samples $n_L$ displays an approximately linear
dependence on $V_g$. A fit of the form $n_L = C_L(V_g - V_{gL}^0)$
gives $C_L \approx 55.7\times10^{10}$ cm$^{-2}$V$^{-1}$ and
$23\times10^{10}$ cm$^{-2}$V$^{-1}$, and $V_{gL}^0 \approx
-0.918\pm0.003$ V and $-0.806\pm0.005$ V for A07 and A78
respectively.

The calibration of $n_s$ with $V_g$ now allows us to identify an
unexpected correspondence between $n_L$ and $n_s$ for both
samples. Firstly, the slopes $C_s$ and $C_L$ agree with each other
as well as the expected specific capacitance $C_0$. Secondly, the
pinch-off voltages $V_{gL}^0$ and $V_{gs}^0$ are also equal within
the experimental accuracy. To confirm this correspondence, we have
plotted $n_L$ and $n_s$ together as functions of $\Delta V_g$,
where $\Delta V_g = V_g - V_{gL}^0$ for $n_L$ and $V_g - V_{gs}^0$
for $n_s$. Fig.~4b clearly shows $n_L$ to be a continuation of
$n_s$ in the low-density, strongly-localized regime in both
samples. While the best alignment could be obtained using $A_L =
0.0664$, Fig.~4b also illustrates the position of $n_L$ relative
to $n_s$ for A07 with $A_L = 0.143$ (dotted line) and $A_L =
0.083$ (dashed line).

The correspondence between $n_L$ and $n_s$ clearly demonstrates a
distribution of localized states that is independent of the
topology of disorder, but depends only on the density of
electrons. Such a possibility has been discussed extensively in
the context of Wigner crystallization in low-density systems.
Recent quantum Monte Carlo simulations have indicated the
possibility of stabilizing a Wigner-like rigid distribution of
electrons at relatively high electron densities in presence of
optimum disorder~\cite{stc}. Microwave absorption studies have
also claimed possible observation of Wigner solid in the
fractional quantum Hall regime at $n_s$ as high as $\sim
6\times10^{10}$ cm$^{-2}$~\cite{pdye}. Transport in such systems
has been discussed in terms of hopping of vacancies or
interstitials, or the excitation of such species to the mobility
edge~\cite{okamoto,cock}. In this experiment, even though we have
probed the spatial distribution of localized states in a more
direct manner, we cannot, at least at present, make a definite
statement on any order in the distribution. This would require
further understanding of the geometrical parameters, in particular
$A_L$, which is presently under investigation.

The temperature dependence of $\alpha$ provides further insight to
the nature of the ``impurity'' band and corresponding DOS ($D_L$).
In the inset of Fig.~3b we have shown a typical $T$$-$dependence
of $\alpha$, corresponding to $V_g = -0.760$ V for A78. Over the
observed range of $T$, $\alpha$ is nearly constant. In the
percolation framework, this confirms the validity of
Eq.~\ref{alpha} for our system and allows us to express the
bandwidth $W$ as $W \lesssim k_BT\ln{(\eta_c)}$, where $\eta_c$
determines the resistivity at the percolation threshold, $\rho =
\rho_0\exp{(\eta_c)}$. To get an estimate of $W$, we use the
observed $\rho/\rho_0 \approx 6$ at $V_g = -0.76$ V and $T = 0.3$
K. ($\rho_0 \sim h/e^2$ was obtained from $T$$-$dependence of
$\rho$.) This gives $W \lesssim 0.05$ meV, and from the observed
$n_L \approx 1\times10^{10}$ cm$^{-2}$ at the same $V_g$, we get
$D_L \sim n_L/W \approx 2.1\times10^{14}$ cm$^{-2}$eV$^{-1}$.
Interestingly, this is nearly an order of magnitude larger than
even the spin-degenerate free-electron DOS in GaAs,
$m^*/\pi\hbar^2 = 2.7\times10^{13}$ cm$^{-2}$eV$^{-1}$. Similar
observation was reported in the context of transport near the
mobility edge in the inversion layer of Si-MOSFETs~\cite{mp}.
While a shift in the mobility edge as a function of $E_F$ was
suggested, in our case possibility of an interaction-induced
enhancement of the effective mass ($m^*$) cannot be ruled out.

An important aspect of the solid phase is the spin of individual
localized states. Rigid electronic distribution at $n_L$ as large
as $1-5\times10^{10}$ cm$^{-2}$ will definitely involve strong
exchange interaction. Both ferromagnetic and anti-ferromagnetic
ground states have been predicted depending on the nature of ring
exchange in this phase~\cite{okamoto,roger}. Indeed, possible
observation of a spontaneously spin-polarized phase, in particular
at lower dimensions, is now being reported in several experimental
investigation in the low-density regime~\cite{okamoto,ghosh}.

In conclusion, magneto-transport in the strongly localized,
low-density regime of high-quality 2D electron systems provide
evidence of a dynamic, gate voltage-dependent distribution of
localized states. Over a mesoscopic region of the 2D systems, the
spatial density of the localized states was found to agree with
the estimated density of electrons at the same gate voltage. This
correspondence is argued to be manifestation of a many-body rigid
distribution of electrons, driven by strong Coulomb interaction.

\begin{figure*}[h]
\includegraphics[height=14cm,width=18.2cm]{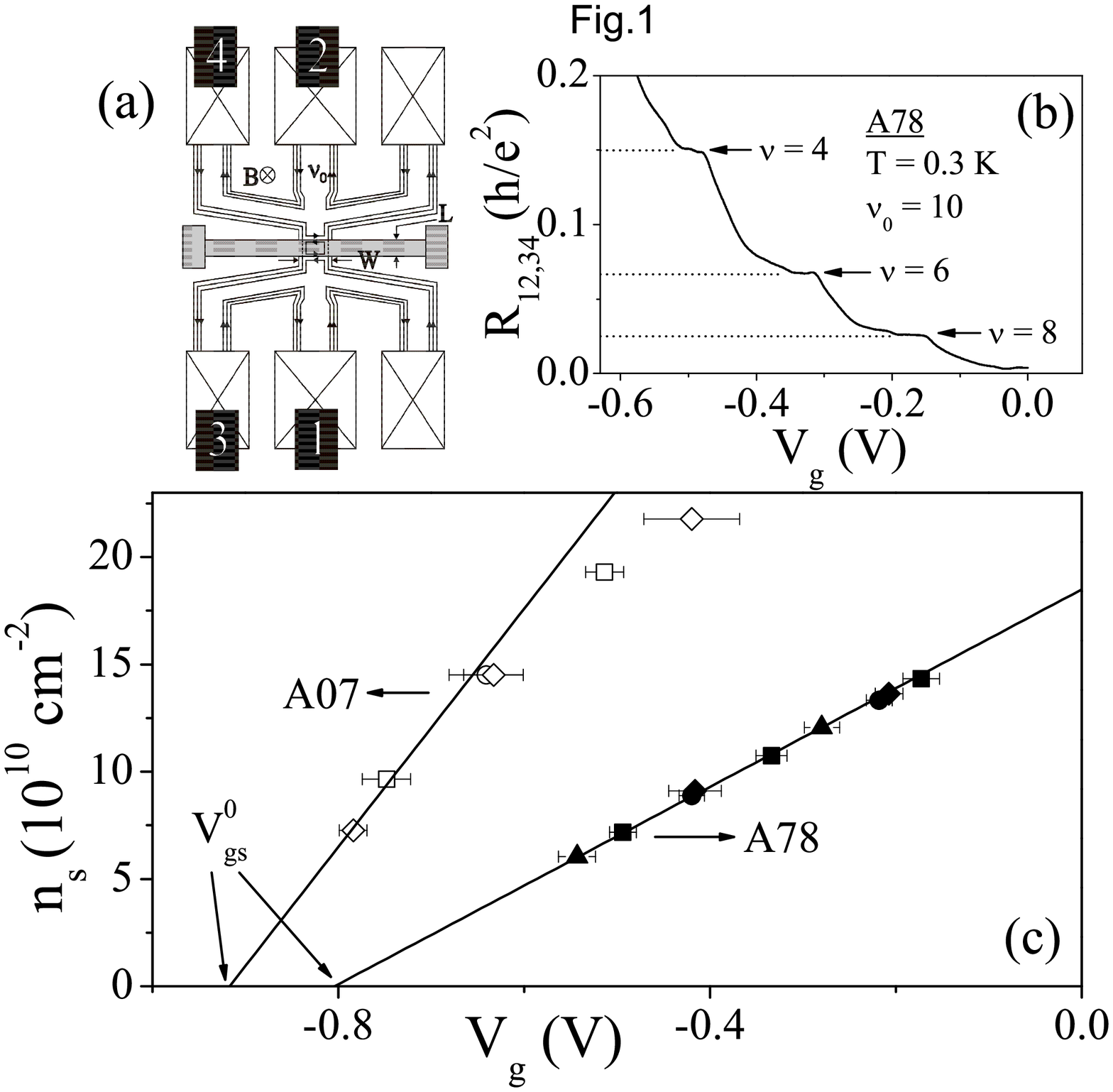}
\caption{(a) A schematic of the device structure. Contacts 1 and 2
are for current probes, while 3 and 4 used for voltage probes. (b)
Illustration of gate voltage ($V_g$) dependence of the 4-probe
resistance $R_{12,34}$ at a constant magnetic field ($B = 0.742$
T) corresponding to an integral filling factor $\nu_0$ (= 10) in
sample A78. The dashed lines are the expected values of
$R_{12,34}$ at different filling factors $\nu$ within the gated
region (see text). (c) $n_s = eB\nu/h$, at values of $V_g$'s
corresponding to the center of plateaus in $R_{12,34}$. The
horizontal error bars represent the width of the plateaus in
$V_g$. The solid lines are fits to the linear part of the data.
For both samples, different symbols correspond to different
integral value of $\nu_0$ at which $R_{12,34}-V_g$ sweeps were
recorded.}
\end{figure*}
\begin{figure*}[h]
\includegraphics[height=14cm,width=18.2cm]{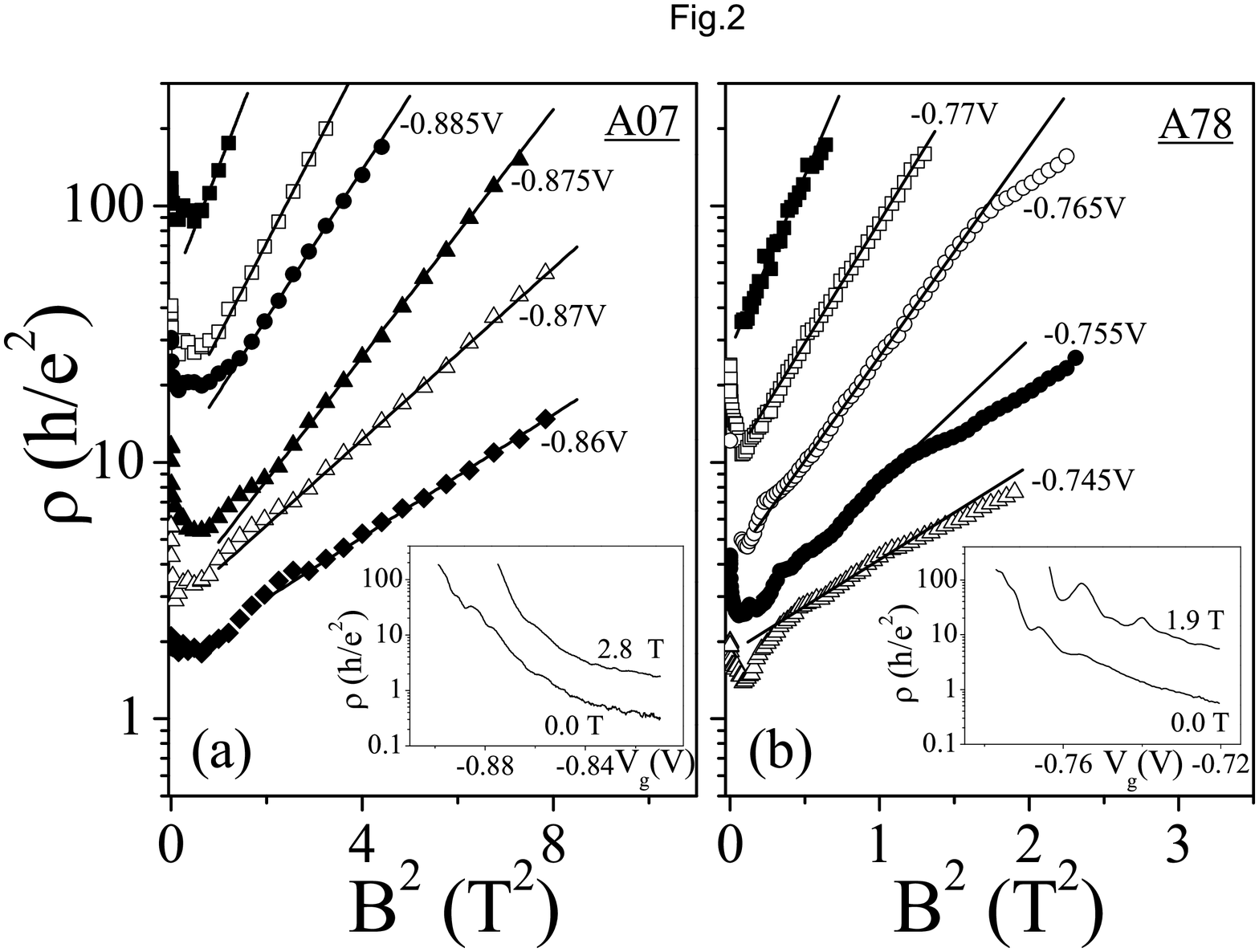}
\caption{Dependence of resistivity $\rho$ on $B^2$ for different
values of gate voltage ($V_g$) recorded at $T = 0.3$ K. Inset:
$V_g$$-$dependence of $\rho$ at zero and finite $B$. (a) Sample
A07, and (b) Sample A78.}
\end{figure*}
\begin{figure*}[h]
\includegraphics[height=14cm,width=18.2cm]{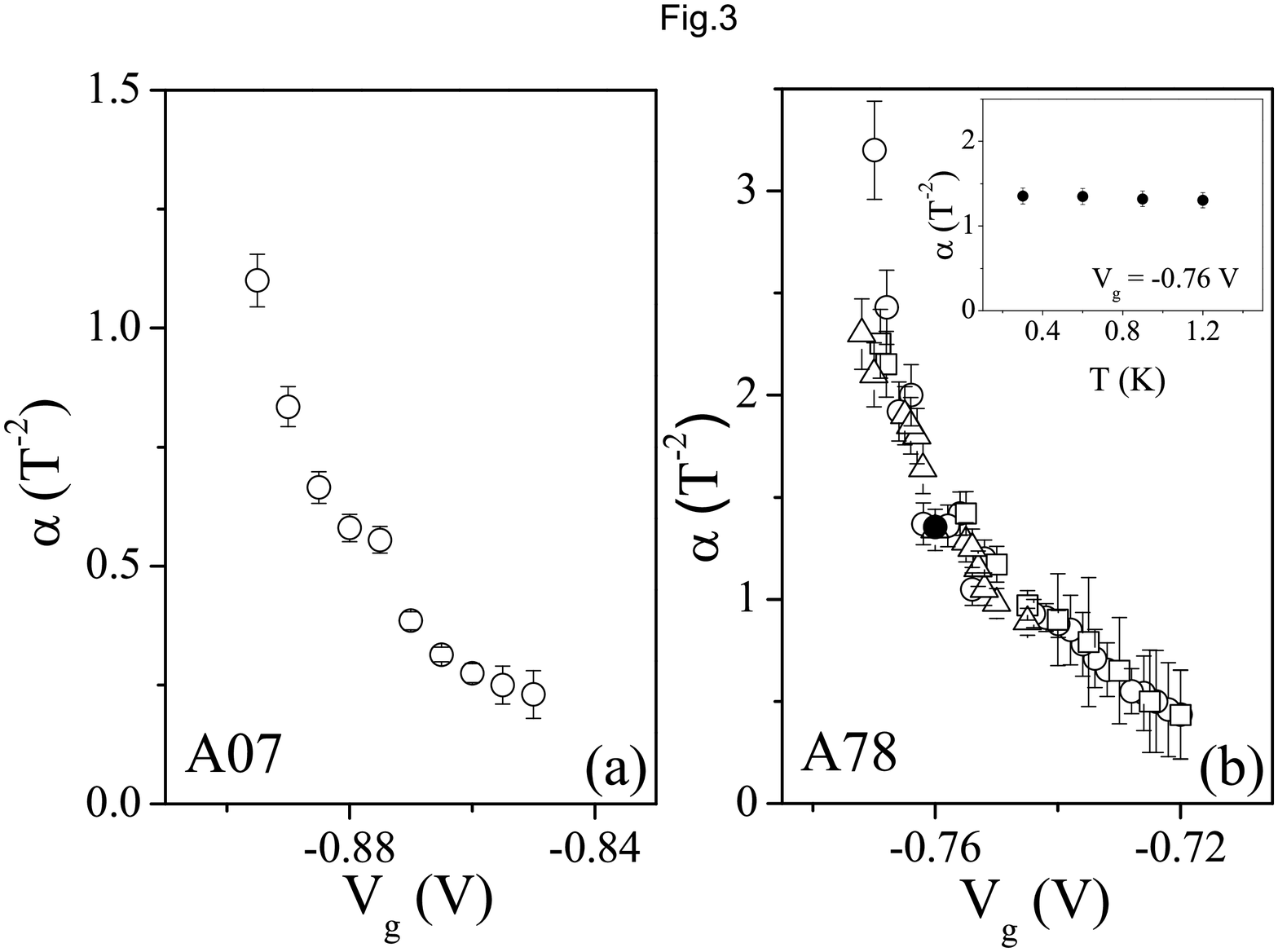}
\caption{Gate voltage ($V_g$) dependence of the slope ($\alpha$)
of the linear region of $\ln{(\rho)}-B^2$ data shown in Fig.~2.
The data was obtained at $T = 0.3$ K. The fit uncertainty
increases, and hence the error bar on $\alpha$, as $\rho$
decreases to $\sim h/e^2$. (a) Sample A07, and (b) Sample A78.
Inset of (b): the typical temperature dependence of $\alpha$
obtained at $V_g = -0.760$ V for A78.}
\end{figure*}
\begin{figure*}[h]
\includegraphics[height=14cm,width=18.2cm]{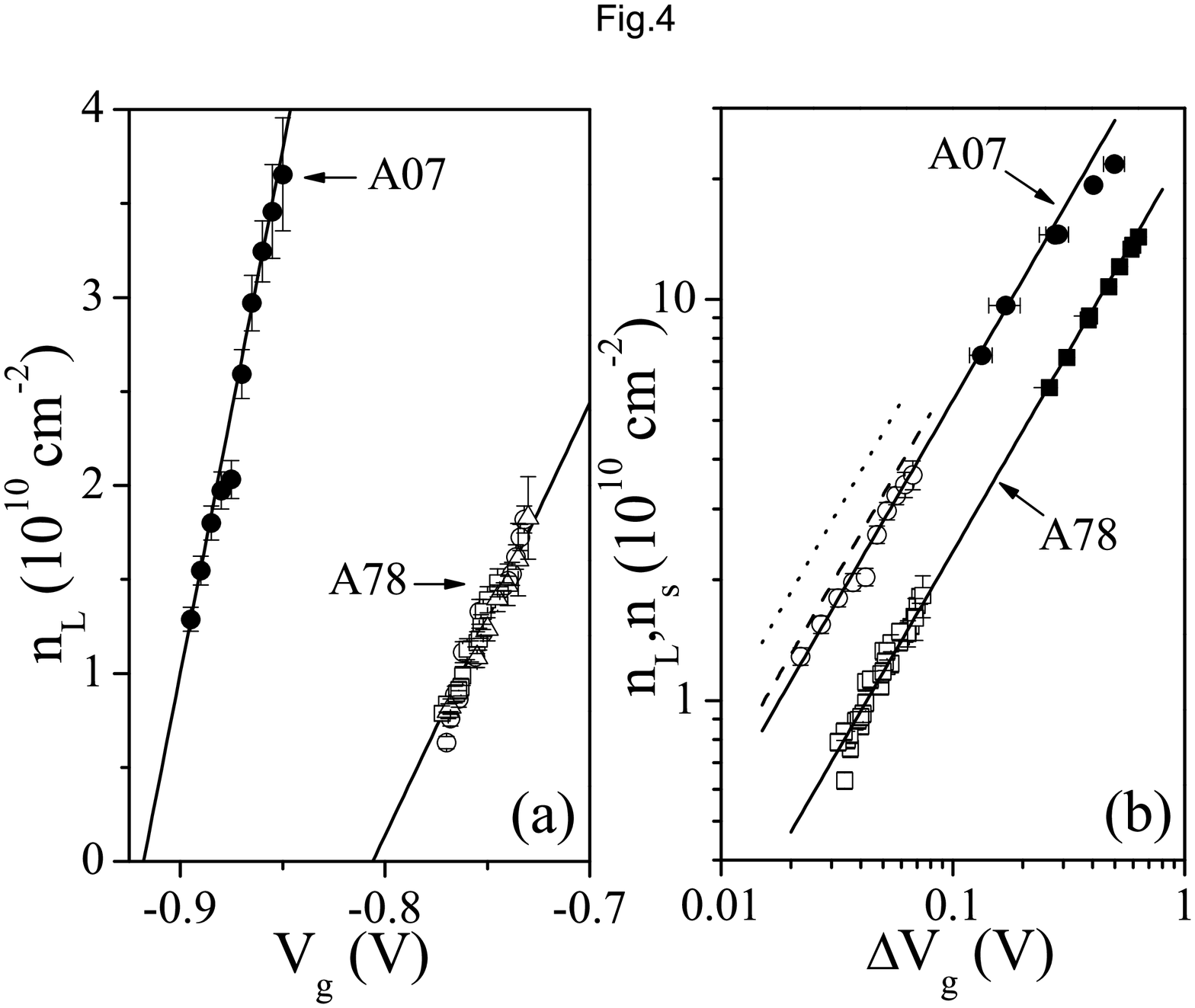}
\caption{(a) The spatial density of localized states ($n_L$) as a
function of $V_g$ calculated from Eq.~1 and the experimentally
obtained $\alpha$. Solid lines denote linear fit. (b)
Correspondence between $n_L$ and $n_s$. For comparison, $n_L$,
$n_s$ $-$ $\Delta V_g$ are plotted in log-log scale, where $\Delta
V_g = V_g - V_{gL}^0$ for $n_L$ (empty symbols) and $V_g -
V_{gs}^0$ for $n_s$ (filled symbols). For A07, the solid, dashed
and dotted lines represent estimates of $n_L$ assuming $A_L =
0.0664, 0.083$ and 0.143 respectively.}
\end{figure*}

\end{document}